\documentclass[a4paper, aps, pra, twocolumn, superscriptaddress,accepted=2026-08-26]{quantumarticle}
\pdfoutput=1

\usepackage[pdftex]{graphicx}
\usepackage{hyperref}
\usepackage{mathtools}

\usepackage[utf8]{inputenc}
\usepackage[T1]{fontenc}
\usepackage{braket}
\usepackage{xcolor}

\usepackage{float}
\makeatletter
\let\newfloat\newfloat@ltx
\makeatother

\usepackage[caption=false]{subfig}
\usepackage{ragged2e} \DeclareCaptionJustification{justified}{\justifying}

\newcommand{\CZ}{$\mathrm{CZ}$}
\newcommand{\CNOT}{$\mathrm{CNOT}$}
\newcommand{\Had}{$\mathrm{H}$}
\newcommand{\PZ}{$\mathrm{Z}$}
\newcommand{\PX}{$\mathrm{X}$}
\newcommand{\PY}{$\mathrm{Y}$}
\newcommand{\ber}{\textit{ber}}
\newcommand{\cer}{\textit{cer}}

\begin{document}

\title{Performance of the spin qubit shuttling architecture for a surface code implementation}
\author{Berat Yenilen}
\affiliation{\footnotesize Institute for Quantum Information, RWTH Aachen University, 52074 Aachen, Germany}
\affiliation{\footnotesize Institute for Theoretical Nanoelectronics (PGI-2), Forschungszentrum Jülich, 52428 Jülich, Germany}
\author{Arnau Sala}
\affiliation{JARA-FIT Institute for Quantum Information, Forschungszentrum J\"ulich GmbH and RWTH Aachen University, Aachen, Germany}
\author{Hendrik Bluhm}
\affiliation{JARA-FIT Institute for Quantum Information, Forschungszentrum J\"ulich GmbH and RWTH Aachen University, Aachen, Germany}
\affiliation{ARQUE Systems GmbH, 52074 Aachen, Germany}
\author{Markus Müller}
\affiliation{\footnotesize Institute for Quantum Information, RWTH Aachen University, 52074 Aachen, Germany}
\affiliation{\footnotesize Institute for Theoretical Nanoelectronics (PGI-2), Forschungszentrum Jülich, 52428 Jülich, Germany}
\author{Manuel Rispler}
\affiliation{\footnotesize Institute for Quantum Information, RWTH Aachen University, 52074 Aachen, Germany}
\affiliation{\footnotesize Institute for Theoretical Nanoelectronics (PGI-2), Forschungszentrum Jülich, 52428 Jülich, Germany}
\email{rispler@physik.rwth-aachen.de}
\begin{abstract}
Qubit shuttling promises to advance some quantum computing platforms to the qubit register sizes needed for effective quantum error correction (QEC), but it also introduces additional errors whose impact on QEC code performance must be evaluated.
The established method to investigate the performance of QEC codes in a realistic scenario is to employ a standard noise model known as circuit-level noise, where all quantum operations are modeled as noisy. 
In the present work, we take the circuit-level noise model and single out the effect of shuttling errors by introducing them as an additional so-called error location. 
This hardware abstraction is motivated by the SpinBus architecture and allows a systematic numerical investigation to map out the resulting two-dimensional parameter space of circuit noise and shuttling error.
To this end, we take the Surface code and perform large scale simulations, most notably extracting the threshold across said two-dimensional parameter space.
We study two scenarios for shuttling errors, on the one hand a depolarizing shuttling error and on the other hand a purely dephasing shuttling error. 
For a purely dephasing shuttling error, we find a threshold of several percent, provided that all other operations have a high fidelity. 
The qubit overhead needed to reach a logical error rate of $10^{-12}$ (known as the "teraquop" regime~\cite{Gidney2021Jul}) increases only moderately for shuttling error rates up to about 1 \% per shuttling operation. 
The error rates at which practically useful, i.e. well below threshold error correction is predicted to be possible are comfortably higher than what is expected to be achievable for spin qubits. Our results thus show that it is reasonable to expect shuttling operations to fall below threshold already at surprisingly large error rates. With realistic efforts in the near term, this offers positive prospects for spin qubit based quantum processors as a viable avenue for scalable fault-tolerant error-corrected quantum computing. 
\end{abstract}

\maketitle

\section{Introduction}

Quantum computation offers to efficiently solve certain computational tasks beyond the reach of classical computers~\cite{Shor1995, Dalzell2025}.
While quantum systems are unavoidably plagued by environmental and operational noise, these detrimental effects can be systematically overcome by the tools of quantum error correction (QEC) and fault tolerance~\cite{Terhal2015}. 
The basic principle of QEC is to use many noisy \emph{physical} qubits to encode few \emph{logical} qubits by quantum error correcting codes. 
Once all components implementing the QEC code can be operated below a certain noise level known as the noise \emph{threshold}, the error rates on logical qubits can be arbitrarily well suppressed by increasing the code size~\cite{Aharonov1999, Aliferis2006}. 
Building qubit devices with substantial numbers of physical qubits on which we can run QEC codes, is thus a key experimental challenge. 
While there is a significant focus on developing  Noisy Intermediate-Scale Quantum (NISQ) applications that operate without full quantum error correction, it is widely believed that the most promising and best-understood applications of quantum computing will ultimately require the implementation of robust QEC to be feasible~\cite{Preskill2018}.
A leading candidate for the implementation of practical QEC is Kitaev's surface code, which among the codes implementable in a two-dimensional geometry with local interactions most notably ranks high in terms of its threshold value under realistic noise models~\cite{Kitaev98,Dennis2002,Raussendorf2007}. 
The operation of increasingly larger instances of this code has been recently experimentally demonstrated in a series of works~\cite{Krinner2022May, Marques2022Jan, Acharya2023Feb, google_QEC_below_threshold_2024, Zhao2022Jul}. 

A central contemporary challenge for all hardware platforms is the quest to operate quantum error correction on hundreds of qubits, where we have seen recent advancements in a neutral atoms~\cite{Bluvstein2023} and superconducting qubits~\cite{google_QEC_below_threshold_2024}. 
In the long term, however, for major error corrected quantum algorithms, millions of qubits are needed~\cite{Shor1995, Dalzell2025}.

A promising modular design approach for scalable quantum computing architectures is qubit shuttling. 
Here, in broad terms, the physical movement of qubits establishes long range connections between distant regions of the quantum processor.
The realization of qubit shuttling without disturbing the qubit state, known as coherent shuttling, is at the forefront of ongoing experimental efforts in particular in trapped ions~\cite{Monroe2021,Kaushal2020Feb,Pino2021, Jain2024}, neutral atoms~\cite{Bluvstein2022,Bluvstein2023} and semiconductor spin qubits~\cite{Vandersypen2017,Langrock2023,Struck2024,Volmer2024,DeSmet2025}. 
The latter has led to the proposal of a concrete shuttling based spin-qubit architecture~\cite{Kunne2024}. 
Shuttling thus adds a new element to the set of qubit operations, inspiring new ideas such as the implementation of effectively three-dimensional qubit lattices in a two-dimensional architecture~\cite{Cai2023}. Also, first works have started to theoretically explore near term spin qubit device capabilities including the smallest error-detecting surface code~\cite{Paraskevopoulos2025}. 
Importantly, shuttling also adds an additional error source whose impact needs to be understood. 
While theory models~\cite{Losert2024, Nagai2025} and first experiments~\cite{DeSmet2025} indicate that shuttling error rates around 0.1\% are conceivable for spin-based platforms, they have not yet been shown consistently in multi-qubit devices 
so that any alleviation of the requirements for error correction can be of major significance.

The goal of the present work is to investigate the impact of shuttling on the threshold and the noise suppression scaling of Kitaev's surface code.
In the QEC literature, these key figures of merit have been established using standard noise models, most notably that of circuit-level noise~\cite{Raussendorf2007, Stephens2014}. 
The basic operations in this model (known as circuit \emph{locations}) are qubit initialization, single- and two-qubit gates as well as idling locations and qubit readout, which are all modeled as noisy. 
We extend this noise model by adding a shuttling location to investigate the influence of shuttling on the operation of quantum error correction. 
Towards this goal, we define an experimentally motivated error model representing the SpinBus architecture~\cite{Kunne2024}, which we call the \emph{bus noise model}. 
With this error model we explore the threshold landscape of the surface code using a finite-size scaling analysis~\cite{Wang2003}.\\
We investigate biased and unbiased shuttling errors and find that the threshold does not change significantly when the shuttling error is of the same order of magnitude as the rest of the errors for both biased and unbiased cases.
Furthermore, at small circuit error rates ($0.1\%)$, the threshold bus error rates are high ($\sim 1.5\%$ and $\sim 4.5\%$ for unbiased and biased cases, respectively), where bus error rates refer to the shuttling error rates and circuit error rates the rest of the errors.
As an additional recently suggested and more hands-on metric, we also report the teraquop qubit count, which is the number of physical qubits required to reach a logical error rate of $10^{-12}$~\cite{Gidney2021Jul}. As a reference point, at a circuit error rate of $0.1\%$ the teraquop qubit count overhead is $\sim 5\times 10^3$ physical qubits per logical qubit without any shuttling bus errors.
We find that turning on the bus error rate at fixed circuit error rate $0.1\%$ does not significantly increase the teraquop qubit count overhead for bus error rates up to $\sim 1\%$ and $\sim 0.3\%$ for biased and unbiased cases, respectively.
Especially in light of the expectation that shuttling errors are mainly dephasing in the SpinBus architecture~\cite{Langrock2023,Losert2024}, our results show that shuttling offers a realistic pathway for the implementation of surface codes.

In the rest of this work, we first discuss the surface code and its decoding processes in Section~\ref{sec:surface_code}, followed by an introduction to qubit shuttling and the SpinBus architecture in Section~\ref{sec:shuttling}. 
We then present the details of the bus noise model in Section~\ref{Section:BEM} and how the key performance metrics (threshold and qubit overhead) are extracted from simulated data in Section~\ref{sec:numerics}.
Finally, we discuss our results in Section \ref{sec:results} and conclude the work with a summary of findings and future implications in Section~\ref{sec:conclusion}.

\section{Surface code}
\label{sec:surface_code}

Kitaev's topological surface code is a stabilizer QEC code with comparatively low weight and geometrically local check operators. 
Additionally, versatile tools have been developed for logical computation using the surface code. 
Notably, lattice surgery~\cite{Horsman2012} enables the mediation of logical entangling gates via strictly two-dimensional local operations.
This makes the surface code a prime candidate for fault-tolerant quantum computation~\cite{Litinski2019} and recent experimental breakthroughs indicate successful operation of the surface code close to break-even~\cite{Acharya2023Feb, Krinner2022May, Marques2022Jan} and also below threshold~\cite{google_QEC_below_threshold_2024}.
In the remainder of this section we define the surface code and review its basic properties. 
Apart from notation, readers familiar with surface code can skip to the subsequent section.
An accessible extensive review on surface code can be found in~\cite{Fowler2012}.

\begin{figure}[ht!]
    \centering
    \includegraphics[width=0.7\columnwidth]{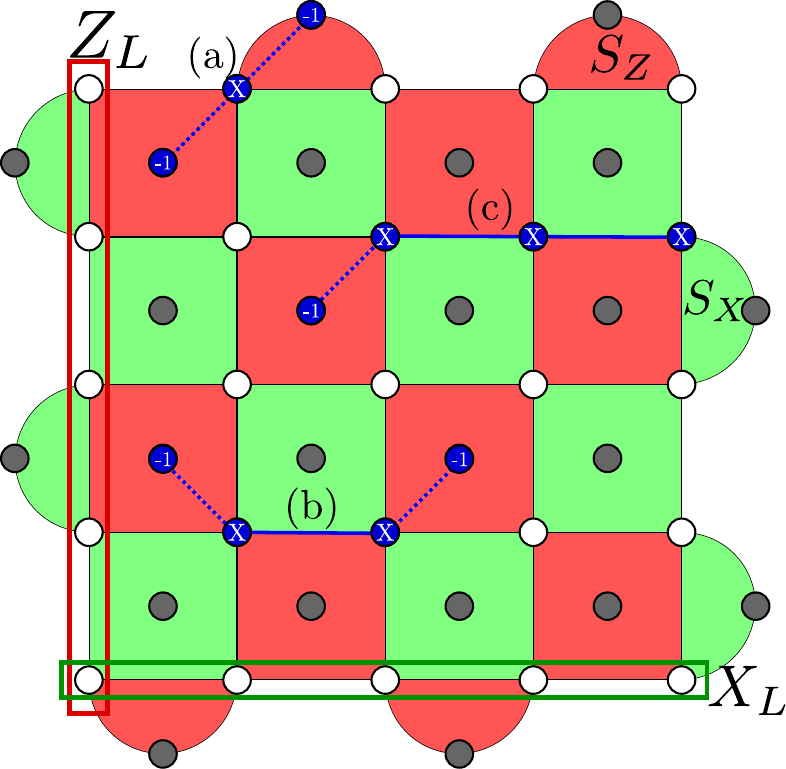}
    \caption{Illustration of a distance five rotated surface code. White (dark-grey) circles represent data (ancilla) qubits. Green (red) plaquettes and half-circles represent \PX{}- (\PZ{}-) stabilizers. Two representatives of logical operators are highlighted by boxes. Three example error chains are represented in (a), (b) and (c) with the underlying errors and stabilizers they trigger. The circuits to measure the stabilizers are shown in Fig.~\ref{fig:stabilizer-circuit}. 
    \label{fig:surface_code_lattice}}
\end{figure}

Departing from Kitaev's original formulation~\cite{Kitaev98, Fowler2012}, we stick to the notion of the rotated surface code~\cite{Horsman2012} where the qubits are identified with the vertices of a square lattice. 
The faces of the square lattice then serve as the stabilizers, alternating between \PX{}- and \PZ{}-stabilizers in a checkerboard pattern as shown in Fig.~\ref{fig:surface_code_lattice}. 
As for any stabilizer code, the code space is defined as the simultaneous $+1$ eigenspace of all stabilizers~\cite{Dennis2002} and measurements of the stabilizers indicate the occurrence of errors on physical qubits. 
For the surface code, the stabilizers at the faces are defined as:
\begin{equation}
    S_X = \bigotimes_{i \in N(g) } X_i, \ \text{and} \quad 
    S_Z = \bigotimes_{j \in N(r) } Z_j 
\end{equation}
where $g$ ($r$) refer to the qubits residing at the green (red) faces, $N(g)$ ($N(r)$) refers to the nearest-neighbors of a qubit residing on a green (red) face, \PX{} and \PZ{} are the conventional Pauli operators.
In addition to these bulk stabilizers, which all form weight-four checks, we implement open boundary conditions by weight-two checks of Z (X) type along the top/bottom (left/right) boundary of the square lattice, indicated by half circles in Fig.~\ref{fig:surface_code_lattice}.

The logical Pauli operators commute with all stabilizers while not being generated from them. 
These are given by \PZ{}- (\PX{}-) Pauli strings extending across the surface vertically (horizontally).
The shortest instances of such operators have length $L$ for an $L\times L$ lattice which means the distance of the code is $L$.
While they can be deformed by multiplication with stabilizers and are thus not unique, they always overlap on an odd number of qubits with each other, such that they mutually anti--commute and thereby satisfy the Pauli algebra on the logical level, defining one logical qubit.

\begin{figure*}[ht!]
    \centering
       \captionsetup[subfigure]{font=large} 

     \subfloat[ ]{
         \includegraphics[width=0.8\columnwidth]{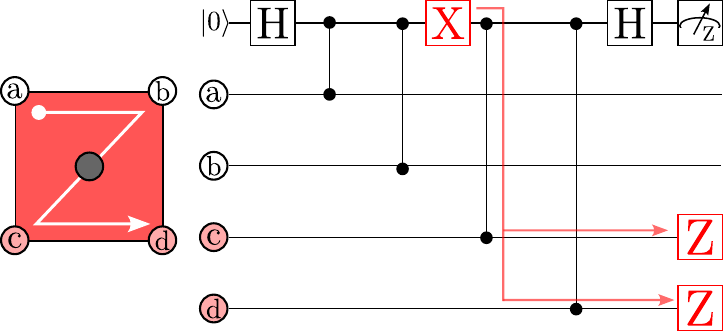}
        }
    \hspace{1.5cm}
     \subfloat[]{
         \includegraphics[width=0.8\columnwidth]{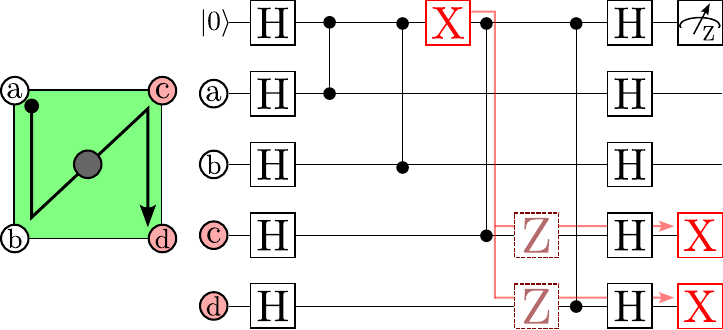} 
        }

    \subfloat[]{
         \includegraphics[width=1.7\columnwidth]{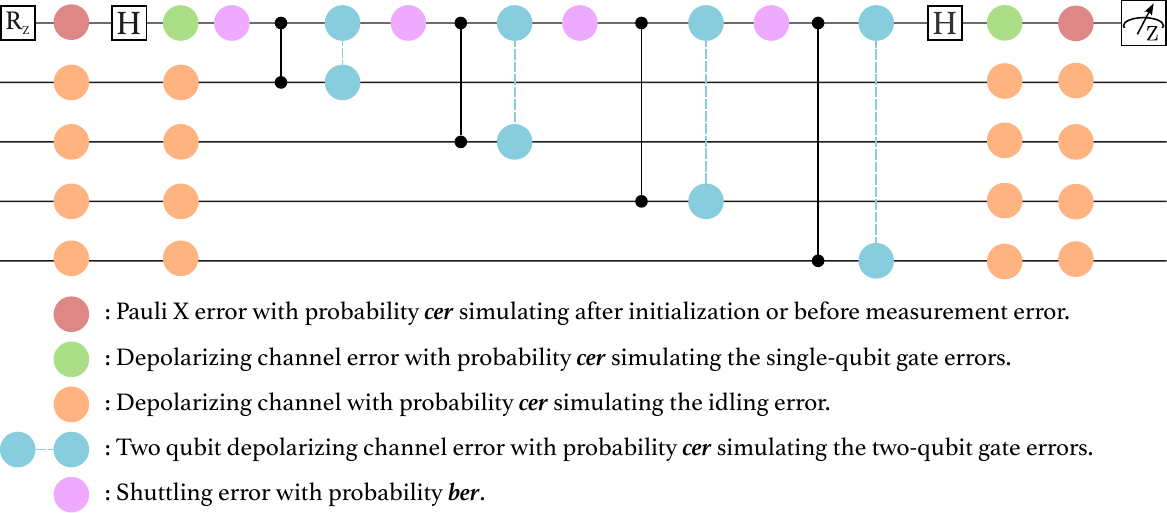} 
     }
     
    \caption{Quantum circuits implementing the measurement of (a)\PZ{}- and (b)\PX{}-stabilizers of the surface code shown in Fig.~\ref{fig:surface_code_lattice}. The given ordering (abcd) of \CZ{} gates allows the \CZ{} gates to be parallelized. Moreover, this choice ensures that the potentially dangerous fault location on the ancilla between the second and third \CZ{} gate propagates onto the code in the direction perpendicular to the respective logical operator, thereby preserving the distance of the code~\cite{Tomita2014}. This dangerous fault location and how it propagates from the ancilla to the data qubits is also illustrated. (c) We define the \emph{bus noise model} (see Sec.~\ref{Section:BEM}), where the standard circuit-level depolarizing noise is extended by additional fault locations on the ancilla qubit, which represents the noise the shuttled ancilla qubit experiences. Exemplarily, we show the Z-stabilizer readout circuit indicating all error locations. Note that for the boundary stabilizer measurements, two of the \CZ{} gates are suppressed such that qubits at the boundary are idle for two time steps. \label{fig:stabilizer-circuit}}
\end{figure*}

\subsection{Syndrome and decoding}
The measurement results obtained from the \mbox{\PX{}- (\PZ{}-)stabilizers} can be used to identify phase-flip (bit-flip) errors on the data qubits. 
Since \PY{}$=i$\PX{}\PZ{}, it is possible to detect and correct all Pauli errors using the syndromes.
This allows us to correct arbitrary single qubit errors on the data qubits.

For example, an \PX{}-error occurring on a single data qubit in the bulk will anti-commute with the two overlapping Z stabilizers which will yield $-1$ syndrome measurements, see Fig.~\ref{fig:surface_code_lattice}(a). 
Chains of data qubit errors will anti-commute with the stabilizers at the two end points of the chain, see Fig.~\ref{fig:surface_code_lattice}(b).
An error chain extending from a boundary into the bulk will trigger only a single syndrome, see Fig.~\ref{fig:surface_code_lattice}(c).

Since errors may in general equally well arise on ancilla qubits, syndromes themselves are rendered unreliable. 
For example, a single bit-flip preceding an ancilla qubit measurement produces the same syndrome as a data error string extending to the boundary.
This entails that from a single round of measurement results we cannot infer which correction to apply to the register.
This issue can be remedied by repeating the stabilizer measurements, recording the changes in syndrome between subsequent rounds, and constructing a \textit{syndrome difference volume}.
This then allows one to distinguish measurement errors and data errors, which on the syndrome difference volume will lead to time-like (ancilla qubit errors) and space-like changes (data qubit errors). 
In order to protect against both types of errors equally well, for an $L \times L$ surface code, it is enough to repeat the stabilizer measurements $L$ times~\cite{Dennis2002}.

Processing the syndrome information in order to come up with a recovery operation is referred to as \emph{decoding}. 
Devising or choosing from a set of known decoding algorithms necessarily entails a trade-off between accuracy and computational efficiency, i.e. maximizing the likelihood of recovery from errors while minimizing the computational resources required to implement said algorithm.
For a recent review on decoding algorithms for the surface code, see~\cite{IOlius2024}.
Minimum-weight perfect matching (MWPM)is a decoding algorithm offering an intuitive understanding and competitive performance, making it a typical choice for surface code decoding~\cite{PyMatchingv2,Fowler2012,google_QEC_below_threshold_2024}. 
Here, non-trivial ($-1$) syndromes are represented as nodes in a complete graph, where the edge weights between any syndrome pair reflect the probability of a corresponding error event. This allows one to translate the  problem of finding the most likely error configuration for a given syndrome to the problem of finding a MWPM, which has an efficient solution~\cite{Edmonds1965}. In this work, we use the recent MWPM implementation PyMatching \cite{PyMatchingv2}.

\section{Qubit shuttling}
\label{sec:shuttling}

Qubit shuttling has emerged as a robust method to overcome geometric limitations in the scalability of several relevant quantum computing systems, enabling flexible long-range connectivity. 
In spin qubit architectures, it can be exploited to reduce cross-talk between nearby qubits during qubit manipulation or while idling. 
It also alleviates the gate-crowding and fan-out problem~\cite{Langrock2023,Kunne2024} that makes dense architectures difficult to wire up by making space for wiring. 
In the longer run, it can create room  for cryogenic control electronics between the primary qubit control elements, which could overcome the limitations of supplying control signals externally ~\cite{Vandersypen2017}. 
Qubit shuttling has been demonstrated in quantum architectures comprised of trapped ions and neutral atoms~\cite{Bluvstein2022,Bluvstein2023, Bowler2012,Pino2021} and of spin qubits~\cite{Baart2016,Fujita2017,Flentje2017,Mills2019,Yoneda2021,Noiri2022,Struck2024}. 
In trapped ions and neutral atoms architectures, the qubits, encoded in the hyperfine sublevels of the trapped atoms, are transported via optical tweezers orfor ions by modulating the DC and AC potentials that create the effective confining potentials~\cite{Monroe2013,Bluvstein2023, Valentini2024}.
In spin qubit architectures, electrons (or holes) are transported adiabatically through a narrow confining channel, and controlled with electrostatic potentials from a series of metallic gates~\cite{Langrock2023}. 
This form of qubit transport has been demonstrated in different modes: (1) the bucket brigade mode~\cite{Mortemousque2021, Zwerver2023, Van_riggelen-doelman2024}, where a qubit tunnels through a series of gate-defined quantum dots and (2) the conveyor mode~\cite{Struck2024, DeSmet2025}, where a smooth, approximately sinusoidal potential moves an electron- (hole-) loaded quantum dot adiabatically.
The model under consideration in this manuscript is motivated by conveyor mode shuttling of electron spin qubits in a Si/SiGe heterostructure.

\subsection{The SpinBus architecture}
\begin{figure}
    \centering
    \includegraphics[width=0.8  \linewidth]{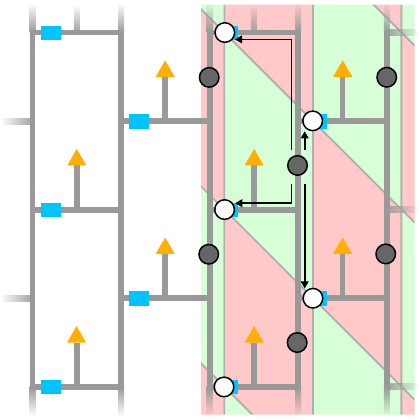}
    \caption{Schematic illustration of the SpinBus architecture. Spin qubits are shuttled in conveyor mode from/to initialization and readout zones (orange triangles), across manipulation zones (blue rectangles) via the shuttling lanes (gray lines). On the right side, the colored background indicates the mapping of the surface code from Fig.~\ref{fig:surface_code_lattice} onto the SpinBus architecture, with white circles indicating the stationary data qubits and grey circles indicating the shuttled ancilla qubits. The arrows indicate the shuttling paths of an ancilla qubit within a green stabilizer. \label{fig:architecture}}
    
\end{figure}

We consider the shuttling-based SpinBus architecture~\cite{Kunne2024}, schematically depicted in Fig.~\ref{fig:architecture}, onto which we map the surface code introduced in Fig~\ref{fig:surface_code_lattice}.
The SpinBus architecture contains initialization and read-out zones (orange triangles), where electrons can be loaded and initialized with the help of a magnetic field gradient created by micromagnets, and where the state of a qubit can likewise be read out; manipulation zones (blue rectangles), where  electric-dipole spin resonance (EDSR) in the presence of a transverse magnetic field gradient enables single- and two-qubit gates, the latter relying on the exchange interaction for state-dependent driving to implement entangling two-qubit gates~\cite{Xue2022,Noiri2022Gate}; and shuttling lanes which are used to transport the electrons across the different elements of the SpinBus architecture, facilitating long-range connectivity with other qubits.

The red-green background of Fig.~\ref{fig:architecture} indicates the mapping of the surface code onto the SpinBus architecture: Stationary data qubits (white circles) are placed near the initialization and manipulation zones, while ancilla qubits (gray circles) can be shuttled between groups of four closest data qubits, as indicated with black arrows. 
In this figure one can see that the periodic structure of the SpinBus architecture and the flexible connectivity provided by the shuttling lanes facilitates an implementation of the surface code.

Shuttling a qubit is a critical operation that must be conducted adiabatically in order to avoid excitations of low lying states---such as valley or orbital states in the case of spin qubits in a Si/SiGe heterostructure---, but fast enough to mitigate the effects of environmental noise~\cite{Struck2024,Ginzel2024}. 
The most prominent expected shuttling errors are the dephasing of the spin qubit due to a virtual or short-lived excitation of a higher valley state,  bit-flips (spin relaxation) at valley hot-spots due to spin-valley hybridization~\cite{Langrock2023,Ginzel2024} as well as nuclear spins and electric noise, similar to stationary qubits. 
In the case of idle qubits---i.e., static spin qubits located in a shuttling lane, away from the stray magnetic fields from the micromagnets in the manipulation and initialization zones---, the probabilities of relaxation and dephasing are $p_\text{r.idl} \sim t/T_1$ and $p_\text{d.idl} \sim (t/T_2)^2$, respectively, with typical relaxation and coherence times $T_1$ and $T_2$ listed in Table~\ref{tab:parameters}. 
We mention in passing that qubit dephasing induced by coupling to the residual nuclear spin background is reduced by the fact that the electron is moving.
This effect known as motional narrowing leads to error rates of~\cite{Langrock2023}.
\begin{align}\label{eq:pdeph}
    p_\text{deph} \approx 2 l_c L_s/(v T_2)^2, \end{align}
where $l_c$ is the correlation length of electric or magnetic noise in conveyor-mode shuttling, $L_s$ is the shuttled length, $v$ is the shuttling velocity. Taking typical parameters from  Table~\ref{tab:parameters}, and assuming a nuclear spin-free landscape, this results in $p_\text{deph}\simeq 10^{-3}$. 

An additional relevant mechanism of decoherence arises from the transport of an electron through a non-uniform valley splitting landscape.
It can cause valley excitations that in turn dephase the spin state via valley-dependent g-factors. 
While a higher velocity reduces the decoherence due to a reduced shuttling duration, it also increases the probability of exciting valley states around spin-valley hotspots. 
This probability can also be kept down to $\sim 10^{-3}$ if the shuttling path is shifted laterally to avoid regions of very small valley splitting and the shuttling velocity is reduced by a factor of $\sim 5$ in their vicinity~\cite{Losert2024}.

Similarly, the probability of relaxation for a shuttled spin qubit is \cite{Langrock2023}
\begin{align}
    p_\text{rel} \approx L_s/(v T_1). \end{align}
Under realistic conditions, dephasing is expected to be the dominant form of decoherence~\cite{Losert2024}, with $p_\text{deph}$ being at least two orders of magnitude larger than $p_\text{rel}$.

Beyond qubit errors, shuttling may also induce leakage to different valley or orbital states, which in general could be detrimental to Surface code operations. The SpinBus design choice of restricting shuttling operations to ancilla qubits is particularly beneficial against leakage: any ancilla qubit leakage will be removed at the end of the stabilizer measurement round following a leakage event. Before removal, a leaked ancilla will lift the exchange coupling such that subsequent pulses now induce random phase errors on the two involved qubits. In total, this converts leakage to regular qubit errors, which subsequent QEC cycles will detect and remove.

We model shuttling errors as acting on the shuttled ancilla only, neglecting data qubit errors due to shuttling crosstalk, as these are expected to be next-order corrections. Conservative estimates for SpinBus-like architectures indicate crosstalk-induced errors of order $10^{-5} - 10^{-4}$ for qubits separated by $\sim 250$nm under resonant driving conditions~\cite{Kunne2024}. Under more realistic off-resonant driving, shuttling is expected to induce even smaller errors on data qubits.

\begin{table}
    \centering
\caption{Estimated typical length and time scales in the SpinBus architecture, as discussed in~\cite{Langrock2023,Ginzel2024,Losert2024}.}
    \label{tab:parameters}
    \begin{tabular}{l|c|c}
        Relaxation time & $T_1$ & 1 s \\ \hline
        Dephasing time & $T_2$ & 20--100 $\mu$s \\ \hline
        1-qubit gate time & $T_{1\text{q}}$ & 100 ns \\ \hline
        2-qubit gate time & $T_{2\text{q}}$ & 50 ns \\ \hline
        Typical shuttling length & $L_s$ & 10 $\mu$m \\  \hline
        Shuttling velocity & $v$ & 2--10 m/s \\ \hline
        Correlation length & $l_c$ & 0.1 $\mu$m \\ \end{tabular}
\end{table}

\section{Bus Noise Model} 
\label{Section:BEM}

It is central for the assumptions that go into our noise model that in the SpinBus architecture only the ancilla qubits of the surface code are shuttled to the interaction zones. This allows us to model the noise from the shuttling process as a Pauli error channel affecting only the shuttled ancilla qubits.
We use depolarizing and Pauli \PZ{}-error channels to model the qubit relaxation and dephasing errors during shuttling, respectively.
We explore two models: an unbiased and a fully biased model.  Defining Pauli noise bias as $\eta = \frac{p_z}{p_x+p_y}~\cite{Tuckett2018Jan}$, this corresponds to $\eta=1/2$ (unbiased) and $\eta=\infty$ (fully biased).
In the unbiased model, we assume no prior knowledge of the error type during shuttling which serves as a baseline.
Therefore, we use a standard depolarizing error channel where Pauli \PX{}-, \PY{}-, or \PZ{}-errors occur with equal probability.
The (fully) biased error model aims to represent the SpinBus architecture where the shuttling has large $T_1$ and comparatively short $T_2$ which translates to a strong bias towards \PZ{}-errors as opposed to \PX{}- and \PY{}-errors.
Thus, we use a Pauli \PZ{}-error channel to model only the dephasing error.

We combine the shuttle errors with the standard circuit-level noise model to obtain the \textbf{bus noise model}.
We assume that available quantum operations are preparation and measurement in the $\{ \ket{0}, \ket{1} \} $ basis, \Had{}, and \CZ{} gates.
All these operations are modeled as perfect operations followed (or preceded) by an error channel with probability \textbf{circuit error rate} (\cer{}).
We assume that the ancilla qubits are shuttled and experience an error with probability \textbf{bus error rate} (\ber{}).
This leads to the following list of fault locations, as illustrated in Fig.~\ref{fig:stabilizer-circuit}c:

\begin{enumerate}
    \item After resetting and before measuring a qubit, an \PX{}-error is applied with probability \cer{}.
    \item After an \Had{} gate, a depolarizing error channel is applied with probability \cer{}.
    \item Before \CZ{} gates,  the ancilla qubit is shuttled, which incurs a depolarizing or dephasing error channel with probability \ber{}.
    \item After a \CZ{} gate,  a two-qubit depolarizing error channel with probability \cer{} affects the involved qubits.
    \item Idle qubits at a time-step experience a depolarizing error channel with probability \cer{}.
\end{enumerate}
As an alternative perspective on this noise model, one could in principle also fuse the shuttling error into the \CZ{} gate error. This would then make this noise channel an asymmetric two-qubit depolarizing noise channel, which would represent the ratio of gate errors and shuttling errors as the bias in this asymmetric depolarizing channel. Note that since our main motivation is to understand the impact of shuttling errors, we abstain from introducing further idling errors on data qubits during the shuttling of ancilla qubits since their magnitude will depend on details of the experimental implementation, such as the shuttling speed, distance between qubits etc. Potential idling dephasing errors can be very well suppressed by up to two orders of magnitude with standard techniques such as dynamical decoupling~\cite{Zwerver2022} with the expected gate induced error budget being negligible with recent improvements~\cite{Wu2025}, such that we do not include an idling error during shuttling.

We thus have a two-parameter noise model with parameters \ber{} and \cer{}.
For the rotated surface code under the standard circuit-level noise (without the shuttling error) with \CZ{} gates, we find the threshold to be $0.00368(3)$, see Appendix~\ref{ap:stabilizers_w_CZ}.
Note that using \CZ{} gates necessitates a $10$-step schedule for one syndrome cycle, compared to an $8$-step cycle when using \CNOT{} gates, leading to a slightly lower threshold compared to $0.00502(1)$ reported in~\cite{Stephens2014}.

\section{Extraction of Performance Metrics}
\label{sec:numerics}

The threshold value of a QEC code under a noise model and a decoding strategy is the physical error rate that separates the error enhancement regime from the error suppression regime in the large system (thermodynamic) limit.
The threshold thus delineates the regime of operation fidelities we have to surpass in order to achieve error suppression. 
The rate of suppression in this regime then typically scales exponentially with system size, allowing arbitrary suppression with moderate increase in qubit overhead~\cite{Dennis2002}. 
In order to accurately determine threshold values, we perform finite-size scaling analysis, which we briefly motivate here. 

Following~\cite{Wang2003, Stephens2014, Watson2015}, we describe the behavior of the logical error rate as a function of physical error rate and the distance of the code. 
Defining $ \xi = | p - p_{th} |^{- \nu_0 } $ as the correlation length, the logical error rate can be written as a function of $L/\xi$ alone. The definition of the correlation length also defines $\nu_0$, which represents the critical exponent of the correlation length~\cite{Chaikin_Lubensky1995}.
Additionally, the logical error rate depends exponentially on the system size~\cite{Dennis2002}:

\begin{equation}
    \label{Eq:PfailvsL}
    \ln P_{fail} \propto - L .
\end{equation}

We thus write \cite{Watson2015, Wang2003}:

\begin{equation}
    \label{Eq:expansatz}
    P_{fail} = A' e^{-a |p - p_{th} |^{\nu_0} L} .
\end{equation}

By performing a quadratic expansion around the threshold we rewrite this relationship as:

\begin{equation}
\label{main:ansatz}
    P_{fail} = A + B (p - p_{th}) L^{1/ \nu_0} + C  (p - p_{th})^2 L^{2/ \nu_0}.
\end{equation}

We can then run simulations of the surface code for varying physical error rates \textit{p} and distances $L$, and calculate the estimator for the logical error rate as:
\begin{equation}
    P_{fail} = N_{error} / N ,
\end{equation}
where $N_{error}$ denotes the number of times the correction operation introduces a logical error, and $N$ is the total number of shots for a given data point. 
After fitting the simulated data points (values for $P_{fail}, p$ and $L$) to Eq.~(\ref{main:ansatz}), we obtain the fit parameters $A,B,C,p_{th}$, and $\nu_0$.

It is possible to use the ansatz in Eq.~(\ref{main:ansatz}) when the fault locations have different error rates \cite{Stephens2014}.
To do so, we use Eq.~(\ref{main:ansatz}) to determine the threshold for the bus noise model as follows. For each fixed \cer{} (chosen below the threshold), we first determine an approximate \ber{} threshold.
We do this by taking the largest \ber{} at which the logical error rate decreases with increasing distance and the smallest \ber{} at which the logical error rate increases with increasing distance.
Afterwards, we take $6$ data points around the approximate threshold and fit the data to Eq.~(\ref{main:ansatz}) by replacing $p$ with \ber{}.

Another important metric to evaluate the performance of a QEC code is the qubit overhead, i.e. the number of physical qubits required to reach a target logical error rate.
We follow the convention introduced in Ref.~\cite{Gidney2021Dec} and set the target logical error rate as $10^{-12}$.
We define \emph{teraquop qubit count} $T(cer, ber)$ as the number of physical qubits required to realize one trillion logical idle gates using $L$ rounds of stabilizer measurements at a fixed \cer{} and \ber{}.
So with $\sim 10^3 \times T$(\cer{}, \ber{})  physical qubits, one can reliably implement a $\sim 10^3 $ logical qubit quantum circuit with $\sim 10^9$ layers \cite{Gidney2021Dec} which would allow the reliable implementation of algorithms for e.g. quantum chemistry simulation or quantum cryptography~\cite{Gidney2021Dec, Dalzell2025}.

We calculate the teraquop qubit count by using the relation in Eq.~(\ref{Eq:PfailvsL}).
That is, we plot the logical error rate (in log-scale) as a function of code distance for fixed \cer{} and \ber{} in the error suppression regime. 
We then make a line fit to the resulting data and extrapolate it to the desired logical error rate to get the corresponding distance, and use the distance to calculate the teraquop qubit count.
An example of such line fit plot is shown in Appendix~\ref{ap:supplementary_figures}.

We follow \cite{Gidney2021Dec} and calculate the error bars on teraquop counts as follows. 
After performing a least squares fit to $\ln P_{fail}$ vs. $L$ data, we find a higher and a lower slope that increases the least squares cost by one.
We use the resulting line fits that overshoot and undershoot the required number of qubits as error bars in our teraquop qubit count figures, see Fig.~\ref{fig:teraquop_waterfall}.

\section{Results and Discussion}
\label{sec:results}
\begin{figure}
\captionsetup[subfigure]{font=normalsize} \subfloat[Unbiased shuttling error]{
\includegraphics[width=0.85\columnwidth]{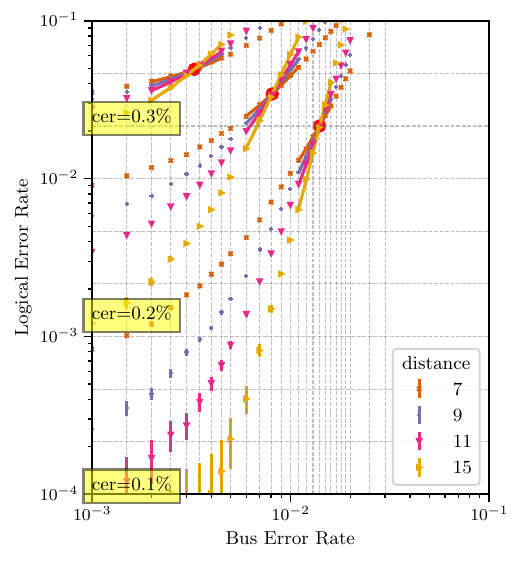}
}\\
\subfloat[Biased shuttling error]{
\includegraphics[width=0.85\columnwidth]{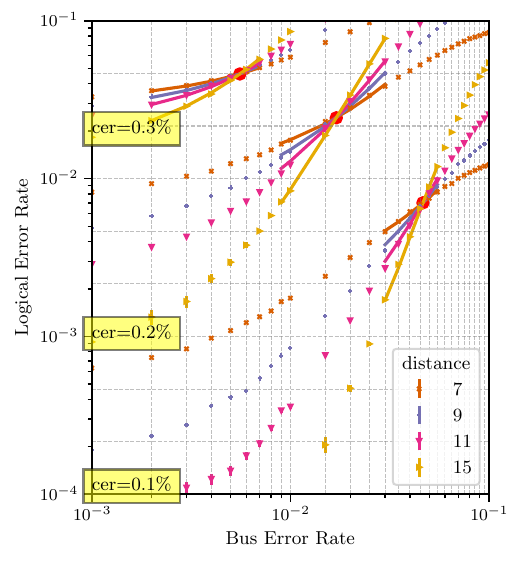} 
}
    \caption{Simulation results for the bus noise model with unbiased (a) and biased (b) shuttling errors near the threshold for distances $\{7,9,11,15\}$.
    The error bars indicate the standard error on the extracted mean for the logical error rate. 
    The red circles specify the threshold point determined by fitting the data for a given \cer{}-slice to Eq.~(\ref{main:ansatz}).
    The resulting fits are plotted around the thresholds.}
    \label{fig:rotated_BNM_p-ber-scan}
\end{figure}

\begin{figure}
    \centering
\includegraphics[width=0.9\columnwidth]{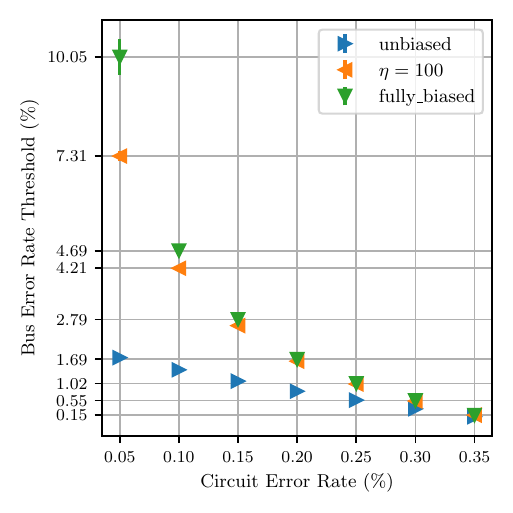}
    \caption{Phase diagram of the surface code under the bus noise model. Data points indicate the Bus Error Rate Threshold value for a given strength of the Circuit Error Rate. Data points (blue: unbiased shuttling noise ($\eta=1/2$), orange: fully biased (pure dephasing) shuttling noise ($\eta=\infty$), green: finite bias ratio $\eta=100$, for definition of $\eta$ see Sec.~\ref{Section:BEM}.) We locate the threshold value from finite size scaling numerical simulations of a given \cer{}-slice to Eq.~(\ref{main:ansatz}). The region between the respective data points and the x-axis corresponds to the below-threshold parameter region where surface code QEC suppresses logical errors arbitrarily well. The error bars indicate the $2\sigma$ region where $\sigma$ is obtained from the resulting covariance matrix.}
    \label{fig:cer-ber}
\end{figure}

We perform simulations for the bus noise model with two different types of shuttling errors with $10^6$ shots for each data point and for odd code distances between 5 and 17.
These results are shown in Fig.~\ref{fig:rotated_BNM_p-ber-scan} for the two noise models defined in Section~\ref{Section:BEM}, the unbiased shuttling error (a) and the biased shuttling error (b).
The data is captured by Eq.~\ref{main:ansatz} around the threshold, which is plotted as a continuous function.
This allows us to extract threshold \ber{} values for each \cer{}-slice using Eq.~(\ref{main:ansatz}), which are indicated as red dots in both figures. 
These threshold values are plotted in Fig.~\ref{fig:cer-ber} as a phase diagram delineating the sub-threshold regime as a function of \cer{} and \ber{}.

A \PZ{}-error on the control (target) qubit before a \CZ{} gate does not propagate to the target (control) qubit and simply add to the ancilla qubit measurement error rate.
Therefore, one may expect the biased shuttle errors to be much more tolerable than the unbiased shuttle errors. 
This expectation is met in our simulations as can be seen in Fig.~\ref{fig:cer-ber} where for \cer{} $< 0.2 \%$, the threshold \ber{} is $\sim 3 - 5$ times more for the biased case than in the unbiased case.
The data thus clearly confirms that the surface code is very tolerant towards pure dephasing (biased) shuttle errors.

Another important point to note in Fig.~\ref{fig:cer-ber} is that even for the unbiased case, shuttling errors do not change the threshold value significantly if \ber{} $\sim$ \cer{} compared to the standard noise model without shuttling.
Furthermore, at small (but plausible\footnote{Note that in this regime idling during shuttling could become a non-negligible contribution, which is not considered here.}) \cer{} $(\sim 0.1 \%)$, the thresholds with respect to \ber{} are up to one full order of magnitude higher compared to the standard circuit-level noise threshold. These findings indicate that errors induced by qubit shuttling do not pose a substantial bottleneck for the implementation of a quantum memory in a surface code.

\begin{figure}
    \centering
    \includegraphics[width=\columnwidth]{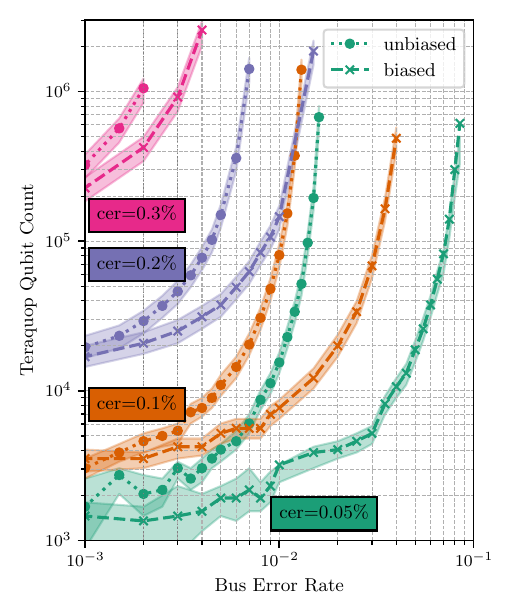}
\caption{Teraquop qubit counts for different circuit error rates. The highlighted regions correspond to over/under shooting the linear fit such that it increases the least squares cost by $1$.}
    \label{fig:teraquop_waterfall}
\end{figure}

To determine the number of qubits required to reach a logical error rate of $10^{-12}$, we plot the teraquop qubit counts vs. bus error rates for the unbiased shuttle error and biased shuttle error in Fig.~\ref{fig:teraquop_waterfall}.
Color map plots for both shuttle errors for all investigated \cer{},\ber{} pairs are presented in Fig.~\ref{fig:rotated_BNM_teraquop_heatmap} in the appendix.

Most notably, for low circuit errors \cer{}$\leq 0.1\%$, bus errors significantly higher than the circuit error rate (up to a remarkable value of about 1\% for biased noise at \cer{}$= 0.05\%$) can be tolerated without increasing the teraquop qubit count by more than a factor two. 
At all error rates, biased bus errors induce less overhead than unbiased ones. 
This difference is largest at small \cer{}. 
Thus, the observations from the threshold regarding the moderate impact of shuttling errors are also reflected by the teraquop study.

\section{Conclusion and Outlook}
\label{sec:conclusion}

In this work, we have studied the effect of qubit shuttling on the performance of surface-code based quantum error correction. 
To this end, we have extended the established realistic noise model of QEC known as circuit-level noise by complementing it with fault locations representing the effects of qubit shuttling, motivated by the hardware layout of a shuttling bus architecture.
This has allowed us to systematically explore the effect of shuttling-induced errors on the overall performance of the surface code.
We have defined two main noise scenarios, on the one hand an unbiased bus noise model, where the shuttling error is modeled as single qubit depolarizing channel and on the other hand a biased bus noise model, where the shuttling error is a purely dephasing channel.
We have performed large scale stabilizer simulations to determine the respective threshold landscapes.
The two resulting phase diagrams constitute a key result of this work and provide the means to understand the interplay between shuttling errors and standard circuit errors when aiming for beneficial operation of quantum error correction.
The foremost general conclusion we draw is that the sensitivity to shuttling errors is substantially less severe that one may have expected from more broad considerations when, e.g., subsuming the shuttling error under circuit operations.
I.e., the threshold is comparatively increased from the general uniform circuit-level noise threshold of $\sim0.5\%$ to up to $\sim5\%$ for the biased bus error.
These results will inform ongoing efforts to enter the era of error-corrected quantum computation.
While the shuttling error is expected to be mostly dephasing dominated, it is a current open question to characterize the noise channel of shuttling in e.g. the bus architecture that motivated our error model.
We leave it for future work to investigate further noise effects along the lines of noise bias, in particular the \CZ{} gate is expected to be dominated by dephasing.
Here it is less clear how to optimally adapt the code and stabilizer readout circuitry to a given bias, since the noise channel will not be contained to the ancilla qubit(s).
It would be interesting to combine our work with previous works that have looked into biased noise, e.g. by deforming the surface code into the so-called $XZZX$ code~\cite{BonillaAtaides2021Apr, Tuckett2018Jan}

As a broader perspective, shuttling will be a vital primitive for the realization of QEC codes beyond the surface code.
Notably, quantum low-density parity-check codes~\cite{qlpdc_review, architecturefastimplementationqldpc} promise to substantially cut the fault tolerance overhead, such as in particular the number of physical qubits required per logical qubit.
There are no-go theorems dictating that these desirable advantages must come at the expense of requiring non-local connectivity~\cite{Bravyi2010}.
Extending our study to such broader classes of QEC codes will be vital for the co-design of quantum hardware and QEC primitives in the quest for universal fault-tolerant quantum computation.

\newpage

\section*{Software}

We use Python~\cite{python} for all coding-related tasks with extensive use of Numpy~\cite{numpy}, Pandas~\cite{pandas}, and Scipy~\cite{scipy} for numerics and Matplotlib~\cite{matplotlib} for visualization; STIM~\cite{Gidney2021Jul} for simulating the surface code stabilizer circuits under the noise models we investigated and PyMatching~\cite{PyMatching} for decoding which uses the \textit{Sparse Blossom algorithm} \cite{PyMatchingv2}.
We also used \cite{gidney_honeycom} as a building block in our code for calculating the teraquop qubit counts and the resulting error intervals.

\section*{Data and code availability}
All the data and the code used to generate the presented results and figures are available in Zenodo \href{https://doi.org/10.5281/zenodo.15011538}{10.5281/zenodo.15011538}. 

\section*{Conflicts of interest}
H.B. is a co-author on several patents and patent applications relating to the SpinBus architecture. 
He is a shareholder of ARQUE Systems GmbH, which has licensed these patents. M.R. was secondarily employed by ARQUE Systems GmbH for a period of time.
\section*{Author contributions}
B.Y. conducted the simulations and generated the plots under supervision of M.R.. M.R., B.Y. and A.S. wrote the manuscript with input from all authors. 
H.B. and M.M. contributed key insights and feedback on both the original motivation and the results. 

\section*{Acknowledgements}
We acknowledge support from the Deutsche Forschungsgemeinschaft (DFG, German Research Foundation) under Germany’s Excellence Strategy Cluster of Excellence Matter and Light for Quantum Computing (ML4Q) EXC 2004/1 390534769.
Furthermore, M.M. acknowledges support by the European Union’s Horizon Europe research and innovation programme under Grant Agreement No. 101114305 (“MILLENION-SGA1” EU Project) and the ERC Starting Grant QNets through Grant No. 804247, and by the Germany ministry of science and education (BMBF) via the VDI within the project IQuAn. 
This research is also part of the Munich Quantum Valley (K-8), which is supported by the Bavarian state government with funds from the Hightech Agenda Bayern Plus.

\newpage
\appendix

\section{Stabilizer Measurement Circuits with CZ-gates}
\label{ap:stabilizers_w_CZ}
As discussed in Section~\ref{sec:surface_code}, it is possible to parallelize the stabilizer measurement circuits such that they can be performed in the same manner for all lattice sizes with the same number of time-steps.
When the available quantum operations are preparation and measurement in $\{ \ket{0}, \ket{1} \}$ basis, \Had{} and \CNOT{} gates, then all the stabilizers can be measured in $8$ time-steps~\cite{Stephens2014}.
Furthermore, by scheduling \CNOT{} gates as shown in Fig.~\ref{fig:stabilizer-circuit} we can prevent dangerous hook-errors to spread to data qubits~\cite{Tomita2014}.

When the \CNOT{} gate is replaced with the \CZ{} gate, however, the number of time-steps required increases to $10$.
We obtain the stabilizer measurements circuits using \CZ{} gates by using the following identity: 
\begin{equation}
    \mathrm{CNOT}(c,t) = \mathrm{H}_t \cdot \mathrm{CZ}(c,t) \cdot \mathrm{H}_t,
\end{equation}
where (\textit{c,t}) means qubit \textit{c} (\textit{t}) is the control (target) qubit., and $\mathrm{H}_t$ denotes a \Had{} gate on qubit \textit{t}.

This substitution transforms a $1$ time-step \CNOT{} gate into a $3$ time-step operation \Had{}, \CZ{}, and \Had{}.
Note that these gates are applied to qubits across the lattice in a nontrivial manner.
This results in a $16$ time-step circuit which has layers: ancilla qubit initialization, $\mathrm{H}^{(1)}, \mathrm{H}^{(2)}, \mathrm{CZ}^{(1)}, \mathrm{H}^{(3)} ,   \mathrm{H}^{(4)}, \mathrm{CZ}^{(2)}, \mathrm{H}^{(5)}, \mathrm{H}^{(6)}, \mathrm{CZ}^{(3)}, \mathrm{H}^{(7)}$, $ \mathrm{H}^{(8)}, \mathrm{CZ}^{(4)}, \mathrm{H}^{(9)}, \mathrm{H}^{(10)}$, ancilla qubit measurements.
We refrain from using qubit numbers for the operations since they are implementation dependant.
The superscripts are used to differentiate layers of \Had{} and \CZ{}.

We can reduce the total number of time-steps to $10$ by combining H layers together in a non-trivial manner which results in the following stabilizer measurement circuit construction:

\begin{enumerate}
    \item Initialize the ancilla qubits in $\ket{0}$.
    \item Apply \Had{} to all ancilla qubits and to all data qubits that interact with \PX{}-stabilizers in \CZ{}$^{(1)}$.
    \item Apply \CZ{}$^{(1)}$.
    \item Apply \Had{} to all data qubits that interacted with an \PX{}-stabilizer in \CZ{}$^{(1)}$ and will interact with an \PZ{}-stabilizer in \CZ{}$^{(2)}$. Apply \Had{} to all data qubits that will interact with an \PX{}-stabilizer in \CZ{}$^{(2)}$ or \CZ{}$^{(3)}$.
    \item Apply \CZ{}$^{(2)}$.    
    \item Apply \CZ{}$^{(3)}$.
    \item Apply \Had{} to all data qubits that interacted with an \PX{}-stabilizer in the \CZ{}$^{(2)}$ or \CZ{}$^{(3)}$ and will interact with an \PZ{}-stabilizer in the \CZ{}$^{(4)}$. Apply \Had{} to all data qubits that interact with an \PX{}-stabilizer in \CZ{}$^{(4)}$ but has not interacted with an \PX{}-stabilizer in  \CZ{}$^{(2)}$ or  \CZ{}$^{(3)}$.
    \item Apply \CZ{}$^{(4)}$.
    \item Apply \Had{} to all ancilla qubits; to all data qubits that interacted with an \PX{}-stabilizer in  \CZ{}$^{(4)}$ ; to all data qubits that interacted with an \PX{}-stabilizer in  \CZ{}$^{(2)}$ or  \CZ{}$^{(3)}$ and did not interact with any stabilizer in  \CZ{}$^{(4)}$.
    \item Measure all the ancilla qubits.
\end{enumerate}

\begin{figure}[ht]
    \centering
    \includegraphics[width=0.8\columnwidth]{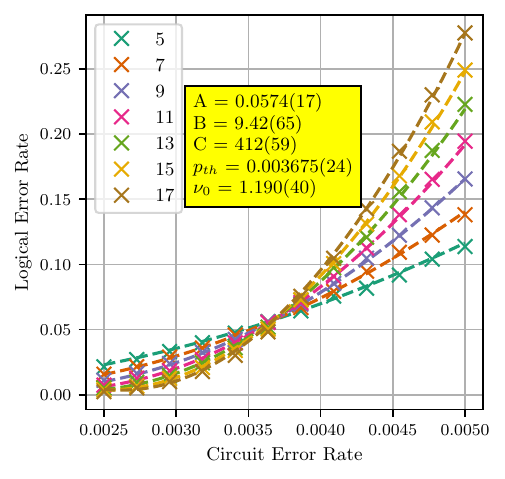}
    \caption{Logical error rate as a function of circuit error rate for the standard circuit-level noise model using \CZ{} gates. $10^6$ shots were used for each data point and the standard error on the mean is used as error bar for each data point (which are plotted but are too small to be visible). The data were fit to Eq.~(\ref{main:ansatz}) and the lines show the fit function. The fit parameters, along with their error bars representing two standard deviations, are also provided. \label{ap:fig:SCBN}}
\end{figure}

In Fig.~\ref{ap:fig:SCBN}, we show the logical error rate as a function of circuit error rate under the standard circuit-level noise model and using the stabilizer circuits with \CZ{} gates which yields a threshold $0.003675(24)$.

\section{Supplementary Figures}
\label{ap:supplementary_figures}
\begin{figure}[]
\captionsetup[subfigure]{font=normalsize} \subfloat[Unbiased shuttling error]{
        \includegraphics[width=0.7\columnwidth]{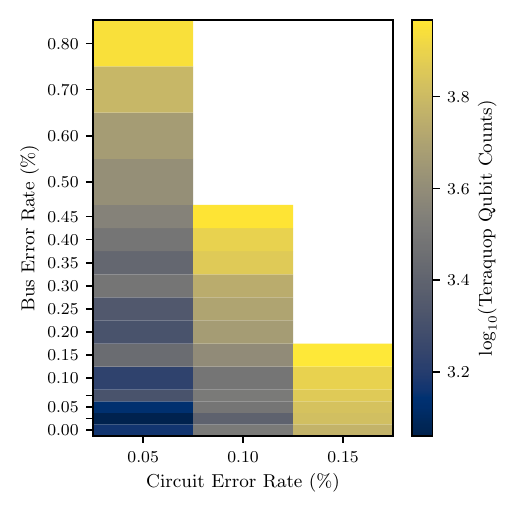}
    }\\
\subfloat[Biased shuttling error]{
        \includegraphics[width=0.7\columnwidth]{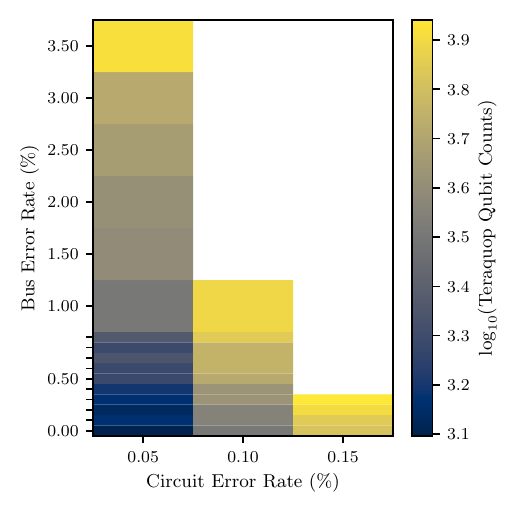}
    }
    \vspace{-0.1cm} \caption{Color map of teraquop qubit counts, see Section~\ref{sec:numerics}A, for (a) unbiased case where the shuttling error is depolarizing and (b) biased case where the shuttling error is only dephasing.}
    \label{fig:rotated_BNM_teraquop_heatmap}
    \end{figure}

\begin{figure*}[t]
    \vspace{0.5cm} \includegraphics[width=1.8\columnwidth]{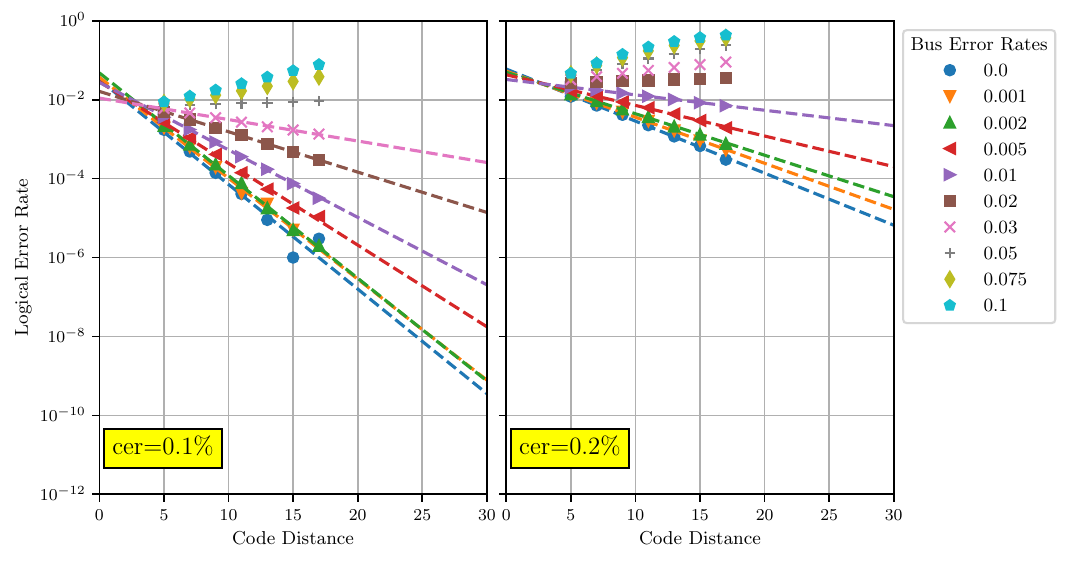}
    \caption{Line fit plots that show logical error rate (in natural logarithm) as a function of the code distance for \cer{} $0.1\%$ (left) and $0.2\%$ (right), and various bus error rates. The fits are performed using least squares fitting and are plotted as dotted lines for the error suppressing cases, i.e. the logical error rate decreases with increasing code distance. The resulting line fits are used to extrapolate the required code distance to reach a target logical error rate.}
    \label{ap:fig:LERSvsDistances}
\end{figure*}

In this section we provide two supplementary figures.
Firstly, we plot the teraquop qubit count for the biased and unbiased bus noise models as a color map of \cer{} and \ber{} which is shown in Fig.~\ref{fig:rotated_BNM_teraquop_heatmap}.
Secondly, we plot $\ln P_{fail}$ (logical error rate in natural logarithm) as function of code distance in Fig.~\ref{ap:fig:LERSvsDistances} for $\cer{} = 0.1\%$ (left) and $\cer{} = 0.2 \%$ (right).
We perform a line fit to the data using the least squares method and these lines are shown for error suppressing cases in Fig.~\ref{ap:fig:LERSvsDistances} as dashed lines.

\clearpage
\bibliography{references}

\end{document}